# Material selection method for a perovskite solar cell design based on the genetic algorithm

Eungkyun Kim
*Electrical and Computer Engineering Department*
*Tennessee Technological University*
Cookeville, USA
ekim43@students.tntech.edu

Indranil Bhattacharya
*Electrical and Computer Engineering Department*
*Tennessee Technological University*
Cookeville, USA
ibhattacharya@tntech.edu

*Abstract*— In this work, we propose a method of selecting the most desirable combinations of material for a perovskite solar cell design utilizing the genetic algorithm. Solar cells based on the methylammonium lead halide, $CH_3NH_3PbX_3$, attract researchers due to the benefits of their high absorption coefficient and sharp Urbach tail, long diffusion length and carrier lifetime, and high carrier mobility. However, their poor stability under exposure to moisture still poses a challenge. In our work, we assigned stability index, power conversion efficiency index, and cost-effectiveness index for each material based on the available experimental data in the literature, and our algorithm determined the $TiO_2/CH_3NH_3PbI_{2.1}Br_{0.9}$/Spiro-OMeTAD as the most well-balanced solution in terms of cost, efficiency, and stability. The proposed method can be extended further to aid the material selection in all-perovskite multijunction solar cell design as more data on perovskite materials become available in the future.

*Keywords—perovskite solar cell, genetic algorithm, HTL, ETL, methylammonium lead halide*

## I. Introduction

Designing high efficiency solar cells while maintaining low cost and high manufacturability is fundamental to commercialization of solar cells. Mono and poly-crystalline Si, quantum dot, organic, and dye-sensitized solar cells do not produce high efficiencies [1]. Multijunction III-V solar cells produce high efficiencies, but their application is limited due to high cost and a difficult manufacturing process. Recently, perovskite based solar cells have attracted wide research interest due to their potential to increase power conversion efficiency and cost-effectiveness, as well as their potential for large scale production. Recent research estimates the cost of a commercial perovskite solar panel to be 66 percent cheaper than a silicon solar panel [2].

While perovskite solar cells have shown success in achieving high power conversion efficiencies, they reportedly suffer from a short lifespan due to their proneness to forming perovskite hydrates in humid environments, leading to the decomposition of the perovskite materials [3]. Additionally, methylammonium lead halide ($CH_3NH_3PbX_3$, X = Cl, Br, or I), the most extensively studied perovskite material for photovoltaic application, has a high lead content, posing concerns around toxicity. The material selection process for perovskite solar cell design must be carefully undertaken to account for the varying degrees of cost-effectiveness, stability, and toxicity of each material. In this work, we propose a method of selecting the best material combinations for a perovskite solar cell design by utilizing the genetic algorithm.

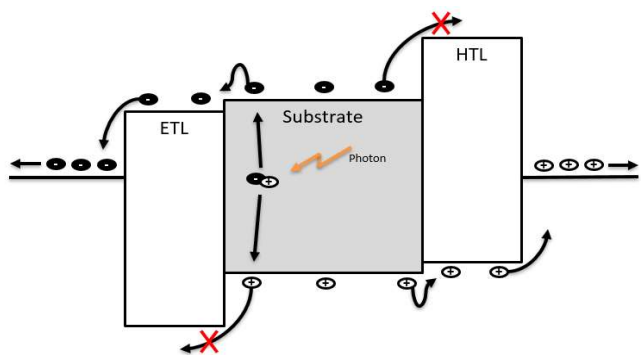

Fig. 1. Energy band diagram of a perovskite solar cell

## II. Structure And Components of Perovskite Solar Cell

A perovskite solar cell is a PIN diode comprised of an active perovskite absorber that is sandwiched between the electron transport layer (ETL) and the hole transport layer (HTL). The fundamental function of an ETL is to form a carrier selective contact with the absorber layer to prevent the holes from diffusing to the counter-electrode and to increase the extraction efficiency of electrons. Similarly, a HTL prevents the electrons from diffusing to the counter-electrode and increases the extraction efficiency of holes. Once the device is under illumination, electron-hole pairs are generated in the perovskite absorber layer, then they are extracted by the ETL/HTL, and finally the carriers are collected by the electrodes, thereby generating photocurrent. The energy band diagram of a typical perovskite solar cell is shown in figure 1.

The perovskite material in the absorption layer plays a core role in photoelectric conversion, and there are many materials of different performance, cost, stability, and toxicity to choose from. Among them, the most widely used material is the methylammonium lead halide, $CH_3NH_3PbX_3$, where X could be either Cl, Br, I, or mixed halides of them. The bandgap of methylammonium lead halide can be controlled by adjusting the *p* orbit of mixed X halide anions, which is especially beneficial in designing a high efficiency perovskite multijunction solar cell [4]. Other perovskite materials used in solar cells include inorganic materials such as $Cs_x(CH_3NH_3)_{1-x}PbI_3$, tin-based lead-free perovskite materials, and many others. While inorganic materials exhibit superior stability and tin-based perovskite materials contain lower levels of toxicity, they do not produce high efficiencies. Therefore, only methylammonium lead halide is considered in this work.

A large number of different HTMs have been explored by researchers. The most widely used HTMs among those are organic HTMs such as spiro-OMeTAD, PEDOT:PSS, PTAA, and P3HT. While organic HTMs exhibit high performance, they suffer from instability and high cost. Inorganic HTMs such as NiO, CuCrO2, Cuo, etc, are inexpensive alternatives with greater stability under heat and moisture. For ETLs, $TiO_2$ has been the most widely studied material due to its wide band gap. Another frequently used inorganic material is ZnO, and organic ETLs such as PCBM have also been developed.

TABLE I : EMPIRICAL DATA ON HTL STABILITY AND DETERMINED INSTABILITY INDEX [5-11]

| HTL Classification | ETL/Perovskite/HTL | η [%] | η aged [%] | Duration / Humidity [hours / %] | Stability index [%] |
|---|---|---|---|---|---|
| Inorganic, Nickel | $TiO_2/CH_3NH_3PbI_3/NiO$ | 12.7 | 12.46 | 300 / 40-60% | 93.06 |
| Inorganic, Nickel | $TiO_2/CH3NH3PbI_3/NiO$ | 10.5 | 10.23 | 300 / 40-60% | 96.18 |
| Inorganic, Nickel | $PBCM/CH_3NH_3PbI_3/NiO$ | 15.71 | 15.10 | 1500 / 30% | 96.53 |
| Inorganic, Copper | $TiO_2/CH_3NH_3PbI_3/Cu-NiO_2$ | 19.62 | 15.6 | 1000 / 35% | 88.54 |
| Organic, Polymer | $TiO_2/CH_3NH_3PbI_3$/spiro-OMeTAD | 13.65 | 3.9 | 600 / 30% | 58.00 |
| Organic, Polymer | $TiO_2/CH_3NH_3PbI_3$/P3HT | 14.4 | 11.9 | 960 / 30% | 91.32 |
| Carbonaceous | $PC_{61}BM/CH_3NH_3PbI_3$/CNO:PEDOT:PSS | 15.26 | 11.2 | 240 / 40 % | 62.5 |
| Carbonaceous | $PC_{61}BM/CH_3NH_3PbI_3$/r-GO | 4.87 | 3.8 | 2000 / 40% | 96.53 |
| Organic, small molecule | $PC_{61}BM /CH_3NH_3PbI_3$/ATT-OMe | 18.13 | 11.8 | 400 / 30% | 59.37 |
| Organic, small molecule | $PC_{61}BM /CH_3NH_3PbI_3$/ATT-OBu | 17.28 | 9.5 | 300 / 30% | 34.72 |
| Organic, small molecule | $PC_{61}BM /CH_3NH_3PbI_3$/ATT-OHe | 15.66 | 8.4 | 400 / 30% | 54 |

TABLE II : BANDGAP ENERGIES OF THE METHYLAMMONIUM LEAD HALIDES

| Material | $E_g$ (eV) |
|---|---|
| $CH_3NH_3PbI_3$ | $1.51 \pm 0.10$ |
| $CH_3NH_3PbBr_3$ | $2.18 \pm 0.10$ |
| $CH_3NH_3PbCl_3$ | $3.02 \pm 0.10$ |
| $CH_3NH_3Pb(I_{1-x}Br_x)_3$ | $(1.51 \sim 2.18) \pm 0.10$ |
| $CH_3NH_3Pb(Br_{1-x}Cl_x)_3$ | $(2.18 \sim 3.02) \pm 0.10$ |
| $CH_3NH_3Pb(I_{1-x}Cl_x)_3$ | $(1.51 \sim 3.02) \pm 0.10$ |

TABLE III: MATERIAL COST OF VARIOUS LEAD HALIDE, ETL, AND HTL MATERIALS, OBTAINED FROM SIGMA-ALDRICH

| Classification | Molecular Formula | Price (USD) |
|---|---|---|
| Lead Halide | $Br_2Pb$ | 10.48/g |
| Lead Halide | $Cl_2Pb$ | 7.76/g |
| Lead Halide | $I_2Pb$ | 3.14/g |
| ETL | $TiO_2$ | 1.6/g |
| ETL | ZnO | 1.25/g |
| ETL | PCBM | 732/g |
| HTL | Spiro-OMeTAD | 507/g |
| HTL | CuSCN | 4.6/mL |
| HTL | P3HT | 484/g |
| HTL | NiO | 4/g |
| HTL | Spiro-OMeTAD | 507/g |

### III. BRIEF OVERVIEW OF THE GENETIC ALGORITHM

With nearly a hundred different materials to choose from, attempting to find the optimum combination of materials by exhaustively fabricating each device quickly becomes expensive and inefficient. By utilizing the genetic algorithm and using the available experimental data in the literature, a desirable combination of materials can be found.

The genetic algorithm is an excellent global search algorithm that easily escapes local maxima. Inspired by the process of natural selection, the genetic algorithm accounts for many ideas such as population, fitness determination, crossover, and mutation. The process begins by creating a random population of N individuals, and each individual is characterized by a set of genes (variables). Once the initial population is created, the fitness scores for all N individuals is calculated. The fitness score is a single figure of merit summarizing an individual's proximity to the optimal solution, and how it is determined is largely dependent upon the application. Based on their fitness scores, a few individuals are selected to be parents whose genes are then combined to generate a new individual (child). This process is referred to as crossover and is repeated until N new individuals are created. Mutation introduces random variation to newly created individuals and is an excellent way to search for a global maximum. A high mutation rate allows the algorithm to escape local traps whereas a low mutation rate allows it to thoroughly search local spaces. The process of fitness calculation, crossover, and mutation continues. After each iteration, individuals with low fitness scores are removed and those with high fitness scores survive to create a new, better generation. The loop continues for either a predefined number of iterations and the best individual is saved or until one or more individuals satisfy the stopping criterion.

### IV. MATERIAL SELECTION BASED ON GENETIC ALGORITHM

Perovskite solar cell individuals are created by first randomly choosing different ETL, HTL, and perovskite materials from the material pools. The HTL and ETL pools contain a variety of organic and inorganic materials, and the perovskite pool contains $CH_3NH_3PbX_3$ where X is Cl, Br, I, or mixed halides of them. There are three indexes associated with each material: stability, efficiency, and cost-effectiveness. Therefore, each perovskite solar cell individual carries 9 distinctive genes: cost-effectiveness, stability, and efficiency index for all HTL, ETL, and a perovskite layer. A population of N perovskite solar cell individuals is shown in figure 2.

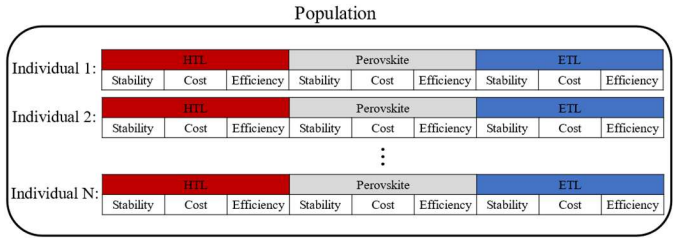

Fig. 2. Population of *N* perovskite solar cell individuals

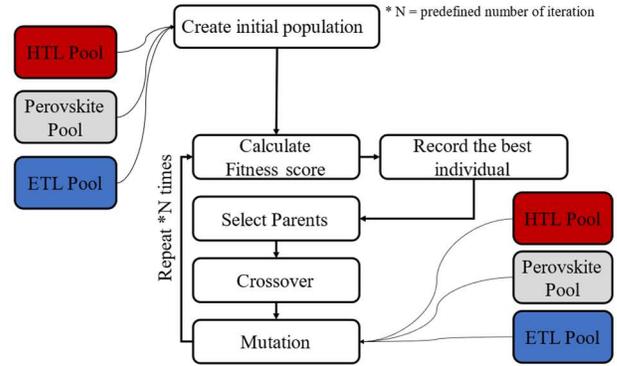

Fig. 3. Flow chart of the material selection process

The stability index of HTL materials is determined based on the available empirical data in the literature. Many experimental research report the power conversion efficiency (η), power conversion efficiency after aging ($\eta_{aged}$), aging duration, and the humidity of the environment under which the device was tested. Assuming that the rate at which the efficiency of a device decreases is linearly proportional to humidity, the stability index is calculated as follows:

$$S = 100\% - \frac{(\eta - \eta_{aged})(100\% - \text{humidity})}{\text{duration}} \quad (1)$$

Once the stability index for all materials are calculated, they are normalized in the range of 0% and 100%. Table I shows the stability index for some of the materials in the HTL pool. It is shown that inorganic materials such as NiO and Cu-NiO$_2$ have a higher stability index, while organic materials have a poorer stability index.

Noh et al reported that $CH_3NH_3PbI_{3-x}Br_x$ with higher Br content (>20%) exhibits higher stability due to the smaller radius or Br, which leads to a more compact structure [12]. Stranks et al. reported that Cl doping improved the stability and achieved good stability with $CH_3NH_3PbI_{3-x}Cl_x$ [13]. This was further supported by Mosconi et al. who reported that the aggregation of Cl ions at the perovskite/ETL interface adjusts the electronic structure of the interface and improves the stability in air [14]. Therefore, $CH_3NH_3PbX_3$ with higher Br and Cl content is given a high stability index. While high Br and Cl content improves the stability, it also increases the bandgap energy, resulting in inefficient absorption of the solar radiation spectrum. For a single junction solar cell, the optimum bandgap energy is 1.34 eV [15]. Therefore, perovskite materials with bandgap energy close to 1.34 eV are given a high efficiency index while those with bandgap energy deviating from 1.34 eV are given a low efficiency index. The bandgap energies of the methylammonium lead halide are listed in table II.

The cost of each material was obtained from a variety of manufacturers' website. Materials with low cost are given a high cost-effectiveness index, and materials with high cost are given a low cost-effectiveness index. Table III lists the price of the most commonly used materials in a perovskite solar cell design.

The fitness score of each individual is then determined by the average of all nine indexes. However, note that depending on the application, this can be easily adjusted by setting different weights on stability, efficiency, and cost-effectiveness. Setting equal weights yields the most balanced solution. Individuals with high fitness scores have higher chance of being selected to be a parent, and parent's genes are then combined with another parent's genes to create a new individual of different material combination. Materials of a newly generated individual then have a certain chance of being replaced by a randomly selected material from the pool based on the predefined mutation rate. This process continues for a predefined number of iterations and the best individual throughout the process is saved. The flow chart of the overall process is shown in figure 3.

## V. Result and Discussion

To obtain the best material combination for a single junction perovskite solar cell, the total population was set to be 10 with a mutation rate of 10%. After 100 iterations, the combination of TiO$_2$ (bandgap 3.2 eV)/$CH_3NH_3PbI_{2.1}Br_{0.9}$ (bandgap 1.96 eV)/Spiro-OMeTAD was found. Although $CH_3NH_3PbI_3$ has a potential for higher efficiency, as its bandgap is closer to the optimum bandgap in the case of a single junction solar cell, our algorithm decided that adding Br content provides a better stability, resulting in a good balance between efficiency and stability. While the stability index of Spiro-OMeTAD is quite low at 58 %, experimentally it has been shown that Spiro-OMeTAD provides superior efficiency compared to other HTLs. Therefore, our algorithm determined that the increase in efficiency exceeds the decrease in stability.

As more data on perovskite materials and various ETL and HTL materials become available in the future, we expect the efficacy of our material selection method to increase significantly. More accurate indexes associated with each material will be attainable, and new indexes such as optical absorptance can be added to the algorithm. For example, once the optical parameters of variety of perovskite materials become available, one can extend this method further to determine the best combination of materials for all-perovskite multijunction solar cells. Photon absorption efficiency over the solar radiation spectrum can be simulated via the transfer matrix method or finite element analysis, and the short circuit current can be obtained by assuming the unity internal quantum efficiency, which can be added to a perovskite solar cell individual as an additional gene.

## VI. Conclusion

In this work, basic structure and carrier transport mechanism of a perovskite solar cell was explained, as well as advantages and disadvantages of some of the commonly used HTL, ETL, and perovskite materials in terms of stability and power conversion efficiency. A brief overview of the genetic algorithm was provided and its application to a perovskite solar cell design was introduced. Our algorithm determined the combination of $TiO_2/CH_3NH_3PbI_{2.1}Br_{0.9}$/Spiro-OMeTAD as the most well-balanced solution in terms of cost, efficiency, and stability. As more data on perovskite materials become available in the future, our algorithm can be easily extended to include new data and improve its performance even further. We believe our material selection method will help to define a path forward to improve the performance of perovskite solar cells.